\documentclass[11pt,a4paper]{article}
\usepackage[utf8]{inputenc}
\usepackage{authblk}
\usepackage{indentfirst}
\usepackage{misccorr}
\usepackage{graphicx}
\usepackage{amsmath}
\usepackage{amssymb}
\usepackage{multicol}
\usepackage{bm}
\usepackage{longtable}
\usepackage{pdflscape}
\usepackage{epstopdf}
\usepackage{multirow}
\usepackage{adjustbox}
\usepackage{amsfonts}
\usepackage{fancyhdr}
\usepackage{ifthen}
\usepackage{fancyheadings}
\usepackage{setspace}

\fancypagestyle{plain}{\chead{\scriptsize \texttt{Communications of BAO, Vol.~65 (Is.~2), 2018, pp.~???-???}}}
\title{Statistical analysis of the new catalogue of CP stars}
\author{S.~Ghazaryan$^{1}$$^{\ast}$, G.~Alecian$^{2}$, A.~A.~Hakobyan$^{1}$}
\affil{\emph{\scriptsize$^{1}$NAS RA V. Ambartsumian Byurakan Astrophysical Observatory (BAO), Armenia\\
\emph{\scriptsize$^{2}$LUTH, CNRS, Observatoire de Paris, PSL University, Universit{\'e} Paris Diderot, 5 Place Jules Janssen, 92190 Meudon, France}\\
\emph{\scriptsize$^{\ast}$E-mail: satenikghazarjan@yahoo.de}}}

\begin{document}
\pagestyle{empty}
\newpage
\pagestyle{fancy}
\label{firstpage}
\date{}
\maketitle

\begin{abstract}
This talk is devoted to the statistical analysis of the new catalogue of Chemically Peculiar stars compiled from papers, where chemical abundances of those stars were given. The catalogue contains chemical abundances and physical parameters of 428 stars based on high-resolution spectroscopy data. Spearman’s rank correlation test was applied for 416 CP (108 HgMn, 188 ApBp and 120 AmFm) stars and the correlation between chemical abundances and different physical parameters (effective temperature, surface gravity and rotational velocity) was checked. From dozens interesting cases we secluded four cases: the Mn peculiarities in HgMn stars, the Ca correlation with respect to effective temperature in AmFm stars, the case of helium and iron in ApBp stars. We applied also Anderson-Darling (AD) test on ApBp stars to check if multiplicity is a determinant parameter for abundance peculiarities.
\end{abstract}


\section{Introduction}

Historically, main sequence Chemically Peculiar (CP) stars were divided in 4 groups by Preston (1974) – AmFm, ApBp, HgMn and $\lambda{Boo}$ stars. Later, two groups of CP stars were added to the known ones - He-weak and He-rich stars. All CP stars have different physical properties. The effective temperatures of all CP stars are in the range of 7000-30000K, some of them are non-magnetic, some- magnetic. More than 50 percent of CP stars are binaries or belongs to the multiple systems. But all those stars have one generality – in the atmospheres of those stars we see peculiarities of different chemical element’ abundances. AmFm stars are non-magnetic and they are characterized by underabundances of calcium and scandium, and high overabundances of iron and nickel (see Kunzli \& North 1998; Gebran, Monier \& Richard 2008, for example). HgMn stars are non-magnetic as well; they are characterized by strong overabundances of mercury and manganese (Hg overabundance riches up to 107) (see Catanzaro, Leone \& Leto 2003; Dolk, Wahlgren \& Hubrig 2003, Alecian et al. 2009, etc.). In contrary, ApBp stars are magnetic and characterized by silicon, chromium, europium and strontium overabundances (see Leckrone 1981; Ryabchikova et al. 1999; Kochukhov et al. 2006, etc.). In this research we studied 3 groups of CP stars – AmFm, ApBp and HgMn stars.

\begin{figure}[t]
\centering
\includegraphics[width=1\textwidth]{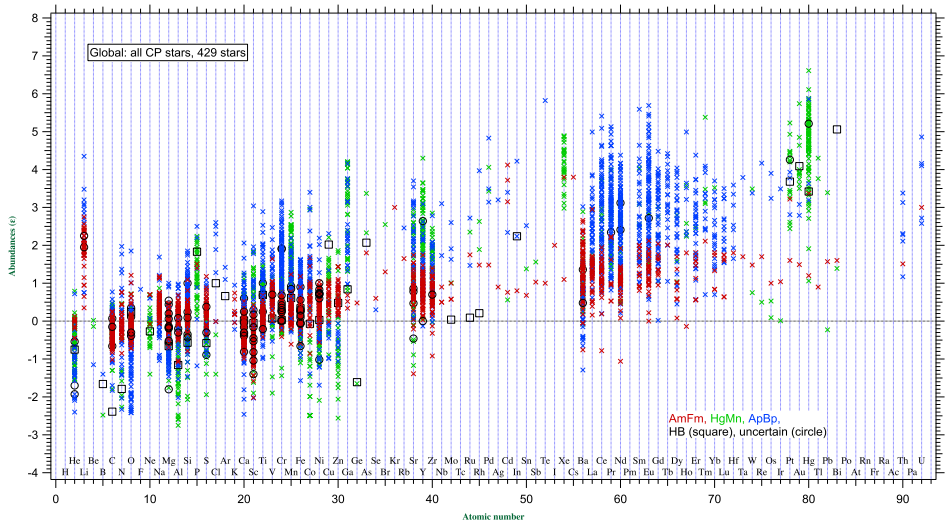}
\caption{Abundances for the three CP Types in the present compilation versus atomic number.
Abundances are the logarithm of the abundances divided by the solar Asplund et al. (2009) ones, the zero line corresponds to solar abundances.
In the HgMn panel, the squares are abundances of the horizontal-branch star Feige 86.}
\end{figure}

\section{The new catalogue of CP stars}

Our new catalogue of CP stars consists in 428 stars, for which all chemical abundances were determined through high resolution spectroscopy techniques. 108 stars are identified as HgMn stars (see Ghazaryan \& Alecian 2016), 128 stars as AmFm and 188 as ApBp stars. The types of 11 stars are uncertain and one star is a known horizontal-branch star (Feige 86). The catalogue contains physical parameters such as effective temperature, gravity, rotational velocity, and chemical abundances with their error measurements. If for a given element the abundances of different ions were given, we took the mean of them for the abundance value, and the error bars were recalculated by the mean square of the errors as in Ghazaryan \& Alecian (2016). Compiled abundances with their errors are shown in Fig.~1. The detailed information on our catalogue is given in Ghazaryan et al. (2018).

\begin{table}[th]
\centering%\begin{center}
\caption{Spearman's rank test results for HgMn stars.
Statistically significant correlations are shown in boldface (\emph{p}-value $\le 0.05$),
marginal ones are underlined ($0.05<$\emph{p}-value $< 0.06$).}
\begin{tabular}{|c|*{9}{c|}}%{|c|c|c|c|c|c|c|c|c|c|}
\hline
%&&$T_{\rm{eff}}$&&&$\log{g}$&&&$v\,\sin{i}$&\\
&\multicolumn{3}{c|}{$\epsilon(T_{\rm{eff}})$}&\multicolumn{3}{c|}{$\epsilon(\log{g})$}&\multicolumn{3}{c|}{$\epsilon(v\,\sin{i})$}\\
\hline
$Elements$&$\rho$&$p$&$N$&$\rho$&$p$&$N$&$\rho$&$p$&$N$\\
\hline
He & \textbf{-0.40} & \textbf{0.001} & \textbf{65} & -0.10 & 0.450 & 65	& 0.12 & 0.355 & 64\\
C & -0.23 & 0.105 & 51 & 0.00 & 0.982 & 51 & 0.03 & 0.841 & 50\\
O & 0.23 & 0.156 & 38 & -0.28 & 0.091 & 38 & 0.15 & 0.391 & 37\\
Mg & \textbf{-0.39} & \textbf{0.001} & \textbf{69} & -0.07 & 0.580 & 69 & 0.09 & 0.489 & 68\\
Al & \textbf{-0.42} & \textbf{0.005} & \textbf{44} & 0.24 & 0.112 & 44 & \textbf{0.32} & \textbf{0.037} & \textbf{44}\\
Si & 0.01 & 0.929 & 74 & -0.15 & 0.192 & 74 & -0.01 & 0.913 & 73\\
S & \textbf{-0.61} & \textbf{0.000} & \textbf{52} & 0.05 & 0.743 & 52 & -0.03 & 0.813 & 51\\
Ti & 0.24 & 0.064 & 62 & 0.11 & 0.403 & 62 & 0.10 & 0.448 & 61\\
Cr & \textbf{-0.28} & \textbf{0.010} & \textbf{86} & -0.02 & 0.828 & 86 & \textbf{0.22} & \textbf{0.041} & \textbf{85}\\
Mn & \textbf{0.48} & \textbf{0.000} & \textbf{70} & 0.06 & 0.650 & 70 & \textbf{0.26} & \textbf{0.032} & \textbf{69}\\
Fe & 0.07 & 0.488 & 90 & 0.06 & 0.561 & 90 & -0.11 & 0.313 & 89\\
Ni & -0.05 & 0.730 & 48 & -0.01 & 0.945 & 48 & 0.11 & 0.473 & 47\\
Cu & 0.32 & 0.106 & 26 & \underline{-0.38} & \underline{0.053} & \underline{26} & 0.22 & 0.289 & 26\\
Zn & -0.18 & 0.339 & 29 & 0.00 & 0.999 & 29 & 0.12 & 0.533 & 29\\
Ga & 0.30 & 0.083 & 34 & 0.23 & 0.184 & 34 & 0.00 & 0.997 & 33\\
Sr & \textbf{-0.49} & \textbf{0.001} & \textbf{45} & \textbf{0.33} & \textbf{0.027} & \textbf{45} & 0.11 & 0.492 & 44\\
Y & -0.17 & 0.220 & 55 & 0.20 & 0.134 & 55 & 0.19 & 0.161 & 54\\
Zr & 0.25 & 0.133 & 37 & 0.25 & 0.136 & 37 & \textbf{0.38} & \textbf{0.021} & \textbf{36}\\
Xe & \textbf{0.41} & \textbf{0.032} & \textbf{28} & -0.04 & 0.823 & 28 & 0.11 & 0.568 & 28\\
Hg & \textbf{-0.22} & \textbf{0.045} & \textbf{86} & \textbf{0.27} & \textbf{0.011} & \textbf{86} & -0.01 & 0.944 & 85\\
\hline
\end{tabular}
%\end{center}
\end{table}

\section{Statistical analysis}

The abundance anomalies may be explained by theoretical models, including atomic diffusion, and for that reason it is interesting to know the correlation between abundances and physical parameters, such as effective temperature, surface gravity and rotational velocity. To check that correlation we applied Spearman's rank test (see Spearman 1904) between abundances and mentioned physical parameters. The statistical analysis was done for each element measured in more than 11 stars.
In Table~1 as an example we show statistical results for HgMn stars. As you see, we found significant correlation with respect to effective temperature for Mg, Al, S, Cr, Mn, Sr, Xe, Hg. Possible correlations suggested by Ghazaryan \& Alecian (2016) for Ni, Ti, and Si are not confirmed in this study. We found also significant correlation with respect to gravity for strontium and confirmed it for mercury suggested in our previous paper. Considering rotation velocity as parameter, correlations have been found for aluminum, chromium, manganese, and zirconium.

Such types of tables were created for ApBp and AmFm stars as well. Dozens correlations between abundances and fundamental parameters were found for those type of stars too. Four noteworthy cases are secluded and explained in detail in our paper (see Ghazaryan et al. 2018).

We have applied AD test on single CP stars and those being in binary systems and do not find any relation between abundance anomalies and multiplicity in all three CP type stars, possibly because of the lack of data.

\section{Conclusions}

We present a unique catalogue of 428 Chemically Peculiar stars observed by spectroscopy during the last decades. Our new compilation of the main physical parameters and the abundances of elements from helium to uranium for 108 HgMn, 188 ApBp, 120 AmFm stars, plus 12 other peculiar stars (including 1 horizontal-branch star) confirms the increase of overabundances for heavy elements with atomic number (see Smith 1996) and the large scatter of the abundances anomalies. This scatter is not only due to the heterogeneity of the data, or abundance determination errors, but it is real.
The applied statistical tests proofs that there is no any correlation between abundance anomalies and multiplicity of HgMn, ApBp and AmFm stars. We have found a significant number of correlations with the effective temperatures, but also some (fewer) with gravity and rotational velocity. We discuss also four noteworthy cases, but this does not mean that there are only four cases that deserve discussion. We are convinced that considering as a whole, the abundance measurements in CP stars will lead to interesting understanding of the physical processes in play in the atmospheres of those stars. In the near future we plan to extend our database to other categories of CP stars (such as stars with helium anomalies), and we have no doubt that a major extension of such a database will be achieved from the final GAIA catalogue.

\section*{\small Acknowledgements}
\scriptsize{This work was supported by the RA MES State Committee of Science, in the frames of the research project N\textsuperscript{\underline{o}} 16YR-1C034.}

\end{document}